\documentclass{pasj00}
\begin{document}
%\SetRunningHead{Author(s) in page-head}{Running Head}
\SetRunningHead{YATSU ET AL.}{INNER RING AROUND PSR B1509$-$58}
%\Received{2008/05/06}%{yyyy/mm/dd}
%\Accepted{2001/01/01}%{yyyy/mm/dd}

\title{Discovery of the Inner Ring around PSR B1509$-$58}

%%% begin:list of authors
% Do NOT capitalize all letters in "textsc".
\author{Yoichi \textsc{Yatsu}, Nobuyuki \textsc{Kawai}}

  \affil{Department of Physics, Tokyo Institute of Technology, 2-12-1
  Ookayama Meguro Tokyo 152-8551}

  \email{yatsu@hp.phys.titech.ac.jp, nkawai@phys.titech.ac.jp}
  
  \author{Shinpei {\textsc Shibata}}
  \affil{Department of Physics, Yamagata University, Yamagata 990-8560}

  \and 
  \author{Wolfgang {\textsc Brinkmann}}

  \affil{Max-Planck-Institut f\"{u}r Extraterrestrische Physik, Postfach
  1603, 85740 Garching, Germany}
%%% end:list of authors
  
%%% Please use the following style in case that sorting by 
%%% affilation is impossible. 
%
% \author{%
%   D-Firstname \textsc{D-Familyname}\altaffilmark{1}
%   E-Firstname \textsc{E-Familyname}\altaffilmark{1,2}
%   and
%   F-Firstname \textsc{F-Familyname}\altaffilmark{2}}
% \altaffiltext{1}{Address of Institute}
% \email{ddddd@xxx.xxx.xx.xx}
% \email{eeeee@xxx.xxx.xx.xx}
% \altaffiltext{2}{Address of Institute}

%% `\KeyWords{}' always has to be placed before `\maketitle'.
 %Do NOT move this preamble from here!

\maketitle

\begin{abstract}
A {\itshape Chandra} study of pulsar wind nebula around the young
energetic pulsar PSR B1509$-$58 is presented.  The high resolution X-ray
image with total exposure time of 190 ks reveals a ring like  feature 10''
apart from the pulsar.  This feature is analogous to the inner ring seen
in the Crab nebula and thus may correspond to a wind termination shock.
The shock radius enables us to constrain the wind magnetization, $\sigma
\geq 0.01$.  The obtained $\sigma$ is one order of magnitude larger than
that of the Crab nebula.  In the pulsar vicinity, the southern jet appears
to extend beyond the wind termination shock, in contrast to the narrow
jet of the Crab.  The revealed morphology of the broad jet is coincident
with the recently proposed theoretical model in which a magnetic hoop
stress diverts and squeezes the post-shock equatorial flow towards the
poloidal direction generating a jet.
\end{abstract}

\section{Introduction}
%Recent understandings
Recent X-ray observations revealed that part of pulsar wind nebulae
(PWNe) have common structures of ``Torus'' and ``Jet'', as typified by
the crab pulsar \citep{2002ApJ...577L..49H,2000ApJ...536L..81W}, the
Vela pulsar \citep{2003ApJ...591.1157P,2001ApJ...556..380H}, and PSR
B1509$-$58 \citep{2002ApJ...569..878G}.  However the physical mechanisms
which form the ubiquitous features, especially the jets, are still
unclear, because the origin of the jet in the Crab is too compact to
discriminate its structure even with the current X-ray observatories.

%Termination shock
The theoretical approaches toward the axisymmetric PWNe have been aimed
to figure out the Crab nebula.  The torus in the Crab had been
successfully explained by one-dimensional MHD simulations
\citep{1984ApJ...283..694K,1984ApJ...283..710K}.  A young rotation
powered pulsar is believed to lose its rotating energy via a magnetized
particle flow from a pulsar, the so-called pulsar wind.  Since the pulsar
wind is flowing almost at the speed of light, it is stalled by a termination shock
just around the pulsar.  When the wind passes through the termination
shock, the kinetic energy of the wind is converted into internal
energy.
%Compression and radiation
Essentially, the post-shock flow is decelerated yielding to the
conservation law of particle flux $nvr^2 = \mathrm{const}$, where $n$ is
the number density (that is almost constant), $v$ is the flow velocity
and $r$ is the radius from the pulsar.  At the same time, the frozen-in
condition, $rvB = \mathrm{const}$, induces the compression of the toroidal
magnetic field $B$, in which the internal energy is efficiently
converted into synchrotron radiation forming an X-ray torus.

%Magnetization control
In practice the magnetic pressure prevents the deceleration of the
post-shock flow, thus the magnetization of the pulsar wind is closely
related to the geometric structure of the PWN.  In this way, several
methods to calculate the magnetization parameter $\sigma$ have been
proposed
\citep{1974MNRAS.167....1R,1984ApJ...283..694K,1994ApJ...435..230G}.
Moreover, recent two-dimensional numerical studies suggested that the
magnetization seems to control the formation of the jet
\citep{2002MNRAS.329L..34L,2004A&A...421.1063D,2006cosp...36..191B}.
Therefore it is essential to evaluate the magnetization parameters to
understand the formation process of PWNe.  

\begin{figure*}
  \begin{center}
    \FigureFile(170 mm,120mm){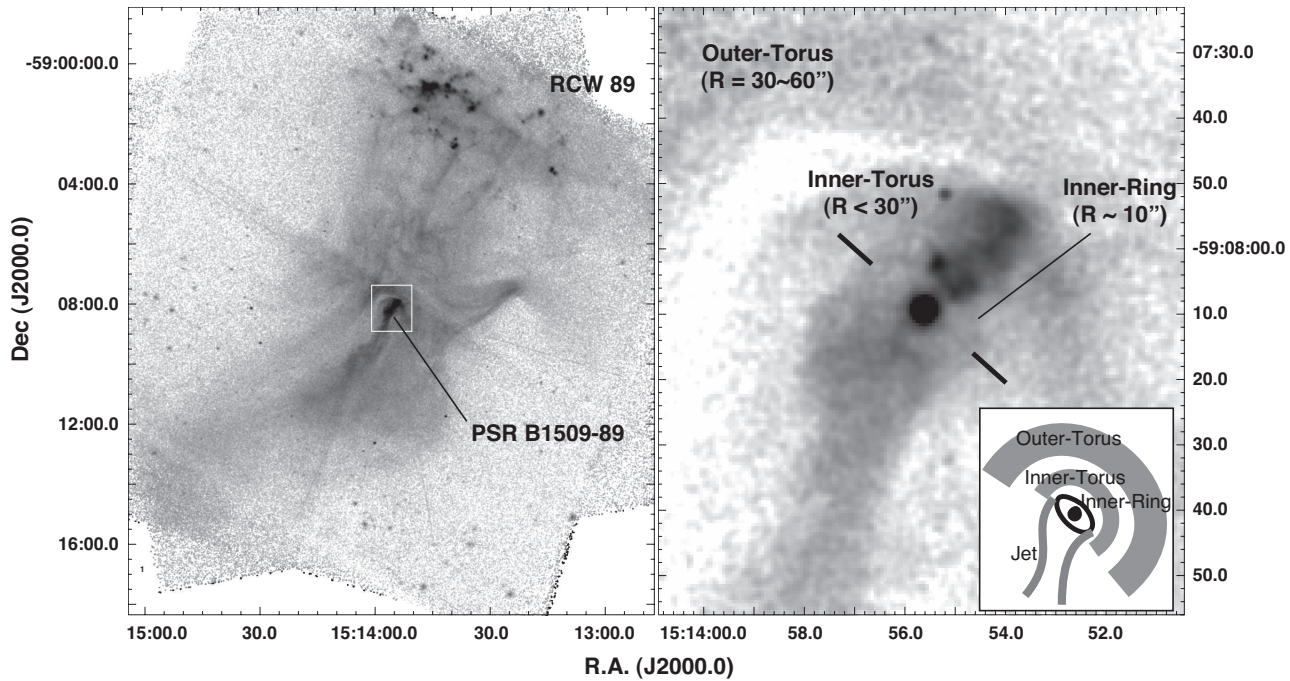}
    %%% \FigureFile(width,height){filename}
  \end{center}
 \caption{(Left)--- Overview image of the PWN accompanied by PSR
 B1509$-$58 with 190 ks exposure taken by {\itshape Chandra}.  The image
 was smoothed with a Gaussian of $\sigma=2.0''$.  The white rectangle
 indicates the FoV of the right panel.  (Right)--- Close-up image of the
 pulsar vicinity smoothed with a Gaussian of $\sigma=0.5''$.  The inner
 panel schematically describes the remarkable structures.}
 \label{X-ray-image}
\end{figure*}

So far, the shock front of the pulsar wind has been observed only in the
Crab nebula \citep{2000ApJ...536L..81W} because of the small angular
sizes of PWNe.  Nevertheless the {\itshape Chandra} X-ray observatory
with sub-arcsec-scale spatial resolution has the potential to detect the
wind termination shocks in other PWNe.  In order to obtain a general
understanding of the formation process of PWNe, we investigate the fine
structures around the pulsar PSR B1509$-$58 utilizing the {\itshape
Chandra} data.

PSR B1509$-$58 is emitting 150 ms radio, X-ray, and Gamma-ray pulsation
\citep{1982ApJ...256L..45S,1982ApJ...262L..31M,1993ApJ...417..738U}.
From the spin parameters, a characteristic age $\tau_\mathrm{c}=1700$
yr, a spin-down luminosity $\dot{E} = 1.8 \times 10^{37}$ ergs s$^{-1}$,
and a surface magnetic field $B_\mathrm{p}=1.5\times10^{13}$ G have been
obtained, making it one of the youngest, the most energetic, and highest
field pulsars known \citep{1982ApJ...256L..45S,1994ApJ...422L..83K}.
The pulsar is accompanied by a bright and large PWN with remarkable
jet-like features extending to the southeast and northwest
\citep{2002ApJ...569..878G}.

The pulsar and its PWN appears to be embedded in a 30 arcmin radio shell
G320.4$-$01.2 \citep{1981MNRAS.195...89C}.  An H{\rm I} observation
yielded the distance of $d=5.2\pm1.4$ kpc from the earth
\citep{1999MNRAS.305..724G}.  Adopting this distance and standard
parameters of the ISM and supernova explosions yields an age of 6-20
kyr, which is one order of magnitude larger than the pulsar's
characteristic age \citep{1983ApJ...267..698S}.

%\citet{2002AJ....123..337D} reported a small number density of $\sim
%0.4$ cm$^{-1}$ based on an H{\rm I} observation.
%THAT PREVIOUS SENTENCE IS UNCLEAR !!!
 Moreover \citet{2005XrU...604..379Y} detected proper motions of X-ray
clumps in RCW 89 which coincide with the north radio shell from the 4.3
yr base-line {\itshape Chandra} observations.  These results imply that
the pulsar and the SNR have the same progenitor in a cavity.  If this is
the case the termination shock of the pulsar wind can be easily
discriminated by {\itshape Chandra}.

The paper is structured as follows.  The {\itshape Chandra} observations
and their results are shown in \S2.  The imaging analysis and spectral
analysis are described in \S3 and \S4, respectively.  Finally, we discuss
the obtained results in \S5.

\section{Observations}
%Observation

\begin{table}[b]
 \begin{center}  
  \label{obssum-1509} \caption{Summary of the observations of PSR
  B1509$-$58.}
  \begin{tabular}{ccccccc}
   \hline\hline
   ObsID\#&Date&Detector&Mode&Exposure\\
   &&&&(s)\\
   \hline
   5534&2004/12/28&ACIS-I&VFaint&50130\\
   5535&2005/02/07&ACIS-I&VFaint&43130\\
   6116&2005/04/29&ACIS-I&VFaint&47650\\
   6117&2005/10/18&ACIS-I&VFaint&46140\\
   \hline
  \end{tabular}
 \end{center}
\end{table}

The vicinity of PSR B1509$-$58 has been observed several times by the {\itshape
Chandra} X-ray Observatory.  For this study, the four most recent
observations which are sequentially performed from 2004 to 2005 are
utilized.  The exposure time of the each observation is about $\sim$ 50
ks.  The aiming point (ACIS-I) was set on (\timeform{15h13m55.60s},
\timeform{-59D08'08.9''}).  The observational settings are summarized in
Table 1. %\ref{obssum-1509}.

%Data reduction
We have performed the reduction process for the each data set using
``acis\_event\_process'' in CIAO version 4.0.  Although these four
observations are performed in VFaint-Mode, we applied standard background
cleaning using $3\times3$ event island, because the VFaint-Mode
background cleaning process tends to misjudge a certain amount of source
photons to be background events.

%Overall structure
Figure \ref{X-ray-image} shows X-ray images combining the four {\itshape
Chandra} observations.  Thanks to the long exposure time of 190 ks,
several fine structures are revealed.  The pulsar wind nebula is very
large, almost filling the radio shell of G320.4$-$01.2.  The length of
the X-ray nebula along the jet-like feature reaches $\sim16'$, which
corresponds to $\sim 24$ pc for the distance $d=5.2$ kpc.  The south
edge of the nebula is more distant ($\sim10'$) from the pulsar than the
north edge ($\sim 6'$),  while the width of the nebula across the
east-west direction is about $8'$, which is a half of the length along
the axis.  

%western cutoff
Interestingly the western side of the PWN appears to be sharply cut off at
$2'$ from the pulsar, implying that the pulsar wind is confined by an
invisible ISM.  Moreover filament-like absorption features are seen at the 
western edge.  Evidently, the column density seems to change at these
features \citep{2008phD...Yatsu.TokyoTech}.  A more detailed treatment and
discussion of these data are deferred to a subsequent paper.

%Compact structures
In the close-up image in Figure \ref{X-ray-image}, nested triplet ring
features, which are labeled ``Outer-Torus'', ``Inner-Torus'', and
``Inner-Ring'', are found.  The outer features have already been 
reported by \citet{2002ApJ...569..878G}.  The outer torus is seen at a
separation of $\sim 60''$ from the pulsar.  In the inner side of the
outer-torus, an inner-torus with a radius of $\sim 30''$ exists.
Most intriguing is an indication of an ``Inner-Ring'' structure $\sim 10''$
apart from the pulsar as seen in the Crab nebula
\citep{2000ApJ...536L..81W}.  Nevertheless the existence of the ring
feature is still ambiguous comparing with the Crab, which might be due to
the contamination from the bright nebula core.

\section{Data Analysis and Results}
\subsection{Imaging analysis }
%Unsharp mask Technique
In order to visualize the compact structures in the vicinity of the
pulsar, we applied an ``unsharp mask'', which is a kind of the high-pass
filter, to the {\itshape Chandra} image.  First, we generated a
low-frequency component image that should be removed.  To avoid the
impact from the bright point sources such as the pulsar, we removed
noticeable point sources using ``wavdetect'' in CIAO 4.0.  The gap
regions were extrapolated from the surrounding areas using ``dmfilth''.
Then the filled-in image was normalized by the exposure map.  Finally,
the exposure corrected image was smoothed using the tool ``aconvolve'' As
a convolution kernel, a Gaussian function of $\sigma=3.5''$ was adopted.
We also generated a source image by the same method as for the
low-frequency image but with a smoothing with a Gaussian of
$\sigma=0.5''$.  Finally, the low-pass filtered component was subtracted
from the source image.

\begin{figure}
  \begin{center}
%    \FigureFile(130 mm,130mm){unsharp_image.ep}s
    \FigureFile(90 mm,90mm){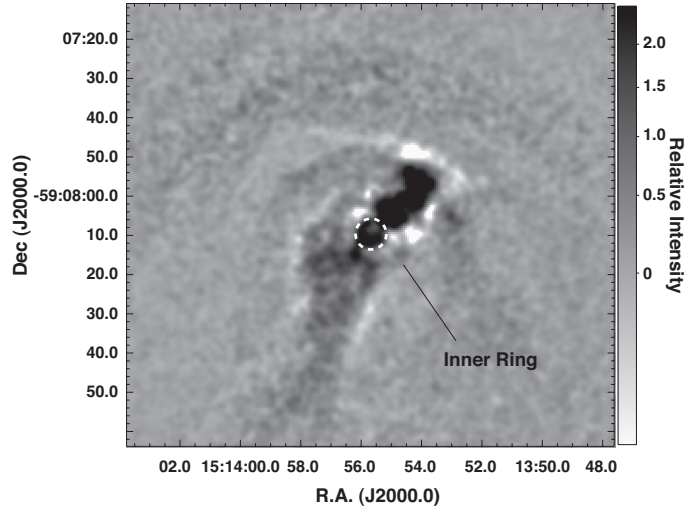}
  \end{center}
 \caption{Unsharp masked image of the pulsar vicinity.  This image was
 generated from a source image smoothed with a Gaussian of
 $\sigma=0.5''$, from which a low-pass image smoothed with a Gaussian of
 $\sigma=3.5''$ was subtracted.  The pulsar's location is indicated by
 the white dashed-line.}  \label{unsharp-masked-image}
\end{figure}

%Resultant image
The resultant image is shown in Figure \ref{unsharp-masked-image}.  The
inner ring is clearly distinguished from the pulsar.  The radius of the
ring feature appears to be $\sim10''$ along the major axis, that
corresponds to $7.8\times 10^{17}$ cm for the distance of 5.2 kpc from
earth.  The eccentricity of $\epsilon \sim 0.8$, for a minor axis
radius of $\sim 6''$, implies an inclination angle of $i\sim50^{\circ}$
which is somewhat smaller than that from the appearance of the outer
torus \citep{2002ApJ...569..878G}.  On the north edge of the inner ring,
there are emitting blobs, which forms the north jet.  These small
structures are seen only in the north side of the pulsar,  while the
south jet appears to be diffuse.  The south jet across the flow line is
about 20'' at the origin, which just corresponds to the radius of the
ring feature.  These structures seems to changes temporally
\citep{2008phD...Yatsu.TokyoTech}.  The detailed temporal analysis will
be reported in a subsequent paper.

\begin{figure}
  \begin{center}
%    \FigureFile(120 mm,120 mm){regions.eps}
    \FigureFile(87 mm,87 mm){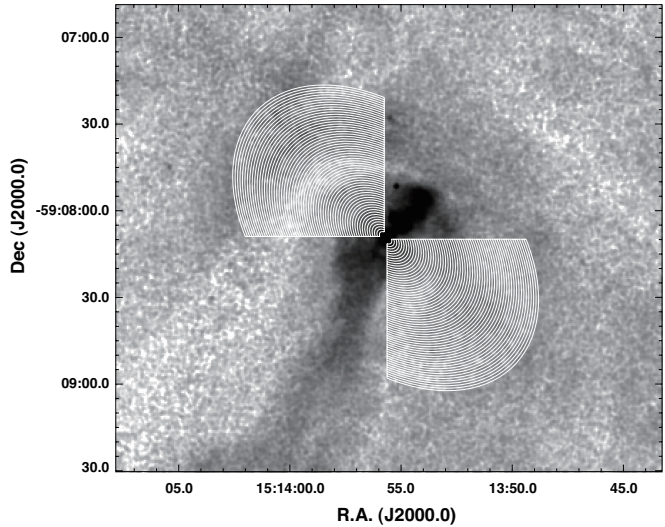}
  \end{center}
 \caption{The sub-regions to be analyzed are described by white lines.
 The each sub-region has a width of $1''.0$ along the major axis
 (North-east$\sim$South-west).} \label{region}
\end{figure}

%%%%%%%%%%%%%%%%%%%%
%Radial profile
%%%%%%%%%%%%%%%%%%%%
%region selection
Next we measure the radius of the ring feature based on the radial
profile of the surface brightness for a quantitative discussion.  As
shown in Figure \ref{region}, sixty sub-regions were selected from which
the X-ray photons were accumulated.  Each of the sub-regions has a width
of 1.0''.  The ellipticity of the regions are aligned with the
inner-ring.  In order to exclude the jet components, the north-west
quadrant and the south-east quadrant respect to the pulsar were masked.
Thus the resultant sub-regions exhibit bow-tie shapes.  

\begin{figure}[t]
  \begin{center}
%    \FigureFile(90mm,120mm){radial_profile_fit.eps}
    \FigureFile(87mm,116mm){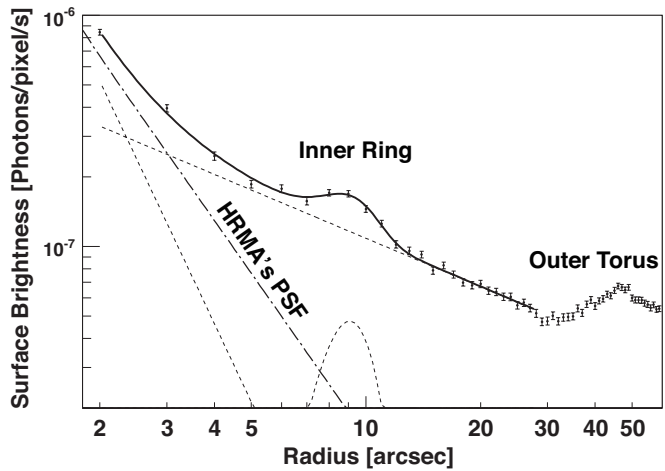}
  \end{center}
 \caption{The radial profile of the X-ray surface brightness at the
 energy band 0.4-8.0 keV.  The solid line shows the model function
 consists of a Gaussian function and two power-law functions.  The
 dashed lines show the each component consisting the model.  The
 dot-dash line shows HRMAs point spread function normalized by the
 pulsar's luminosity at 1'' from the pulsar.}  \label{radial-profile}
\end{figure}

The radial profile of the X-ray surface brightness for 0.4$\sim$8.0 keV
band is shown in Figure \ref{radial-profile}.  The error bars indicate
the statistical uncertainties corresponding the total numbers of the
X-ray photons detected at the each sub-region.  Since a sub-region
accumulates more than 1000 photons, the relative uncertainty is smaller
than $\sim4$ \%.

To constrain the radius of the ring feature, we performed a model
fitting with a multi-component function consisting of two power-law
functions and a Gaussian function.  For the fitting we utilized the data
points ranged in 2''.0 $\sim$ 28''.0.  The centroid of the Gaussian that
corresponds to the inner ring was $9''.0 \pm 0''.1$ at 1 $\sigma$
confidence level.  The other parameters are summarized in Table 
2. %\ref{fit-results}.

\subsection{Spectral analysis}
%Region selection
In this section the spatially resolved spectroscopy is presented.  Since
the photon statistics was a little poor, we investigated the spectral
evolution with a spatial resolution of 3''.  The examined regions are
the north-east side of the bow-tie shape in Figure \ref{region}.

%Model
For the model fittings, we employed a single component model consisting
of an absorbed power-law function.  However, it is difficult to
determine the column density for each region individually, because the
source regions are small.  Therefore we fixed the column density at
$N_\mathrm{H}=1.0\times 10^{22}$ cm$^{-2}$.  With the above
configurations, the four data sets observed in the different epochs could
not be explained by the same parameter set, the energy flux varies with
 time.  These results may be caused by the temporal changes of the
PWN or by the different settings of instrumentation, such as the roll
angle.  Thus we allowed the normalizations of the four data-sets to vary
individually.  

%The model fitting to the X-ray spectrum of the sub-region
%$10''$ from the pulsar is shown in Figure \ref{spectrum}.

%\begin{figure}
%  \begin{center}
%    \FigureFile(90mm,120mm){wisp_ne_03_pegp_nH1.0.eps}
%  \end{center}
% \caption{The X-ray spectrum of the sub-region 10'' from the pulsar.
%% Colors corresponds to the epochs, 2004-Dec-28 (black), 2005-Feb-07 (red),
% 2005-Apr-29 (green), and 2005-Oct-18 (blue).}  \label{spectrum}
%\end{figure}

\begin{figure}
  \begin{center}
    \FigureFile(90mm,120mm){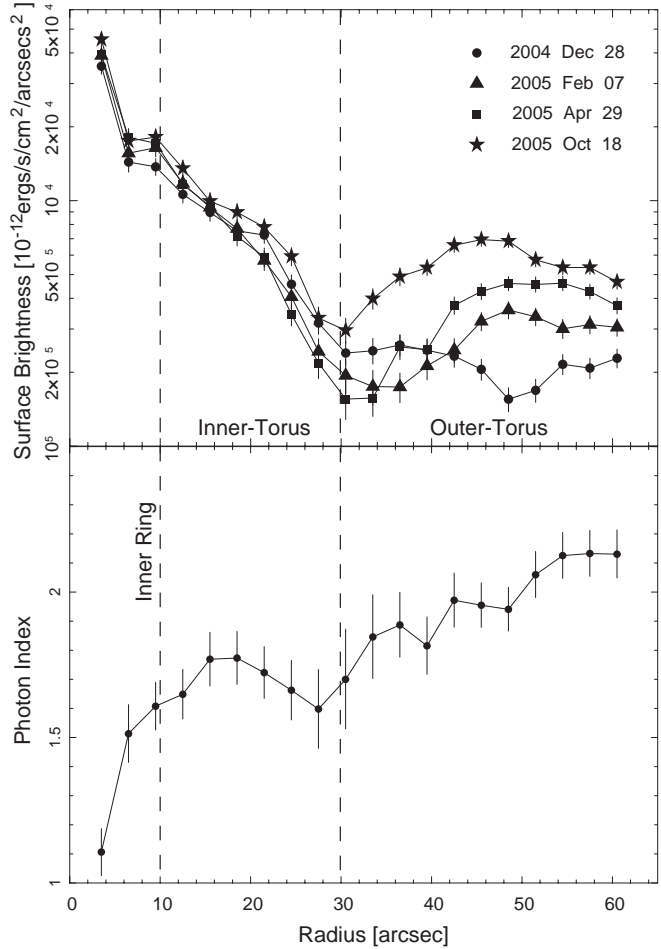}
  \end{center}
 \caption{The evolutions of the surface brightness (upper panel) and the
photon index (lower-panel) are shown as functions of radius.  The shape
of the data points corresponds to the epochs, 2004-Dec-28 (circle),
2005-Feb-07 (triangle), 2005-Apr-29 (square), and 2005-Oct-18 (star).}
\label{params}
\end{figure}

%Results
The results of model fittings are shown in Figure \ref{params},
indicating the evolution of the surface brightness and the photon index
as a function of radius.  In all panels, shapes of the data points
correspond to the epochs, 2004-Dec-28(circle), 2005-Feb-07(triangle),
2005-Apr-29(square), and 2005-Oct-18(star).  The surface brightness
slightly increases at a radius of $\sim 9''$, corresponding the inner
ring.  At $R\sim 30''$, the light curve shows a dip which corresponds to
the gap between the inner torus and the outer-torus.  The photon index
seems to be constant in the inner torus within $R\sim 30''$, while it
increases in the outer-torus outside the radius of $R\sim 30''$.

\begin{table*}[h]
 \begin{center}  
  \label{fit-results}
  \caption{Summary of the model fitting of the radial
  profile.}
  \begin{tabular}{ll}
   \hline\hline
   Parameter.........& Value\\
   \hline
   \multicolumn{2}{c}{\bf Gaussian}\\
   Normalization........
   &$4.76 \pm 0.44$ $(\times 10^{-8}$ photons s$^{-1}$ pixel$^{-1}$)\\
   Mean........
   &$9''.06 \pm 0''.14$\\
   Sigma........
   &$1''.48 \pm  0''.18$\\
   \hline
   \multicolumn{2}{c}{\bf Power-law (1)}\\
   Normalization........
   &$5.36 \pm 0.65 $  ($\times 10^{-7}$ photons s$^{-1}$ pixel$^{-1}$)\\
   Index........
   &$-0.69\pm   0.04$\\
   \hline
   \multicolumn{2}{c}{\bf Power-law (2)}\\
   Normalization........
   &$5.82\pm   1.31$ ($\times 10^{-6}$ photons s$^{-1}$ pixel$^{-1}$)\\
   Index........
   &$-3.48\pm 0.37$\\
    \hline
\footnotetext* The quoted uncertainties are 1 $\sigma$.
  \end{tabular}
 \end{center}
\end{table*}

\section{Discussion}

\subsection{Double tori}
As shown in Figure \ref{radial-profile}, the PWN around PSR B1509$-$58 has a
centrally concentrated surface brightness profile, which may be caused by its
comparatively small angular size or the contamination from the bright
jet structure pointing towards us.  For this reason, it is difficult to
compare the surface brightness evolution of PSR B1509$-$58 with that of
the Crab nebula in a simple way.  Nevertheless PSR B1509$-$58 possesses an
apparently different morphology; the nested double ring structure,
inner-torus ($R\sim30''$) and outer-torus ($30''<R<60''$), lying in the
equatorial plane relative to the spin axis.  

%Spectral difference
Spectroscopy, these two tori, calling the inner torus and the outer
torus, seems to be completely different.  The inner-torus has a constant
photon index of $\Gamma \sim 1.6$ which is the same as that just at the
pulsar.  In contrast, the photon index in the outer-torus obviously
increases with the radius from $\Gamma=1.6$ to 2.1.  The observed
interval of the photon index variation is $\Delta \Gamma\sim0.5$, which
implies that the observed spectral evolution is caused by synchrotron
cooling \citep{1962SvA.....6..317K}.

%Crab ?
In case of the Crab nebula, the post-shock photon index is $\Gamma=2.1$
and it does not change within the torus.  According to
\citet{1984ApJ...283..694K}, the evolution of the surface brightness in
the torus can be explained by the compression of the toroidal magnetic
field, which enhances the synchrotron energy loss.  Thus the bright
torus of the Crab simply corresponds to the region in which most of the
X-ray emitting particles radiate their energy and burn off there.
Consequently the photon index outside of the torus is steeper than
$\Gamma=2.1$, and the surface brightness decays rapidly.

In terms of spectroscopy, the counterpart of the X-ray torus in the Crab
is the inner torus rather than the outer torus of PSR B1509$-$58.  While
in the outer-torus the X-ray emitting charged particles have burnt off
and thus the synchrotron cooling break is passing through the observed
energy band (8.0 keV to 0.4 keV).  Moreover, it was reported that the
outer-torus has a radio counter part \citep{2002ApJ...569..878G}.  These
results of the spectroscopy imply that the outer-torus could not be
simply explained by radiative losses as claimed by
\citet{2002ApJ...569..878G}.

\subsection{Inner Ring}
It is essential to know the radius of the pulsar wind termination shock
to understand the mechanism which forms the structure of a PWN.  So far,
the termination shock of the pulsar wind has been observed only in the
Crab nebula.  In this section we discuss the ring feature discovered
near around PSR B1509$-$58 comparing it with the inner ring of the Crab
nebula.

%Appearance
As the pulsar wind is believed to be ejected almost at the light speed,
the free flowing region should not be radiating as seen in the Crab
nebula.  In the case of PSR B1509$-$58 the region within the ring
feature is also dark.  Comparing it with the tori which are strongly
affected by the Lorenz boost, the surface brightness of the ring feature
appears to be uniform, so that we can distinguish the south edge of the
ring.  For the Crab nebula \cite{2000ApJ...536L..81W} pointed out that
the surface brightness of the inner ring is more uniform in azimuth,
indicating that relativistic beaming is less significant in it than in
the torus.  The observed apparent properties can distinguish the ring
from the tori and then the hypothesis that the ring structure represents
the termination shock can be suggested.  By the way, the ring shows an
obvious elliptic shape, implying that the shock front in an ellipsoidal
surface or a ring viewed with inclination rather than a sphere.
Regarding the issue, recent numerical simulations predict a shock front
with an oblate shape for an anisotropic pulsar wind
\citep{2003MNRAS.344L..93K,2004A&A...421.1063D}.

%pressure balance
At the termination shock, pressure balance between the ram pressure of
the pulsar wind and the post-shock plasma is expected, as seen in the
Crab nebula.  Based on the X-ray image \ref{unsharp-masked-image}, we
obtained a radius of the ring structure of $r\sim 9''$, which
corresponds to $7\times 10^{17}$ cm for the distance of 5.2 kpc to the
pulsar.  Assuming that most of the pulsar's rotating energy,
$\dot{E}=1.8\times 10^{37}$ ergs s$^{-1},$ is transferred to the
relativistic pulsar wind at the light cylinder, we find the ram pressure
of the pulsar wind at the ring structure to be

\begin{eqnarray}
 p_\mathrm{*} &=& \frac{\dot{E}}{4 \pi r_\mathrm{TS}^{2}\phi c}\\
 &=& 9.6\times 10^{-11} \phi^{-1} 
  \left(\frac{d}{5.2\;\mathrm{kpc}}\right)^{-2}
  \left(\frac{R}{9''.0}\right)^{-2}
  \quad \mathrm{dyn}\;\mathrm{cm}^{-2},
\end{eqnarray}
where $\phi$ is the fraction of a sphere covered by the wind.

On the other hand, the inner pressure of the nebula is roughly estimated
by assuming equipartition conditions (Reference).  From the spectral
fits we obtained an X-ray flux of $1.32\times 10^{-13}$ ergs s$^{-1}$
cm$^{-2}$ and a photon index of $\Gamma=1.6$ for the region where the
ring feature exists ($R=10''$) and we infer an unabsorbed luminosity
density at 1 keV $L_\mathrm{1\;keV} = 6.6\times10^{14}$ ergs s$^{-1}$
Hz$^{-1}$.  For simplification we assumed the emitting volume to be a
partial spherical shell of radius $R=10''$, thickness $\Delta R=3''$ and
opening angle of $45^{\circ}$, corresponding to $V=2.6\times10^{53}$
cm$^{3}$, which may be an overestimate.  For a filling factor $f$ and a
ratio of ion to electron energy $\mu$, the minimum energy present in the
source responsible for the synchrotron emission $W_\mathrm{total} \sim
2.9(1+\mu)^{4/7} f^{3/7} \times 10^{43}$ ergs and a magnetic field of
$B\sim35(1+\mu)^{2/5}f^{-2/7}\;\mu$G are inferred.  The corresponding
pressure of the relativistic gas is then $p_\mathrm{neb} =
W_\mathrm{total}/3V\sim 3.7(1+\mu)^{3/7}f^{-4/7} \times10^{-11}$ dyn
cm$^{-2}$ \footnote{For a torus geometry with opening angle $15^{\circ}$
corresponding to $f\sim 0.5$, the internal pressure of $p_\mathrm{neb}
\sim 5.6(1+\mu)^{3/7}\times10^{-11}$ dyn cm$^{-2}$ can be obtained.}.
Note that the obtained inner pressure is a lower limit because the
post-shock flow of a weakly magnetized pulsar wind is believed to be in
particle dominant conditions rather than in equipartition
\citep{1984ApJ...283..694K}.  Taking into account this fact,
$p_\mathrm{*}$ and $p_\mathrm{neb}$ are comparable, implying that the
ram pressure of the pulsar wind and the internal pressure of the PWN is
balanced at the ring feature.  It is therefore inferred that the newly
found ring feature is the termination shock of the pulsar wind.

\subsection{Magnetization of the pulsar wind}
Adopting that the discovered ring feature corresponds to the wind
termination shock, the magnetization parameter $\sigma$ can be
estimated.  As the shock transforms the kinetic energy into thermal
energy, $\sigma$ can be rewritten as \citep{1998nspt.conf..457S}
\begin{eqnarray}
 \sigma&=&\frac{B_1^2/4\pi}{[\mathrm{kinetic \; energy\; density}]}\\
 &\sim&\frac{B_1^2/4\pi}{\mathrm{[postshock\;thermal\;energy\;density]}}\\
 &\sim&\frac{B_1^2/4\pi}{B_\mathrm{Eq}^2/4\pi},
\end{eqnarray}
where $B_1$ is the toroidal magnetic field just behind the shock, and
$E_\mathrm{Eq}$ is the equipartition magnetic field after the shock.
Combining this equation with the equation of continuity
($r_\mathrm{TS}^2 c/3 = r_\mathrm{n}^2 v_\mathrm{n}$) and the flux
conservation law ($r_\mathrm{TS}cB_1=r_\mathrm{n}B_\mathrm{Eq}$) yields
\begin{equation}
\frac{r_\mathrm{n}}{r_\mathrm{TS}} \sim \frac{1}{3\sqrt{\sigma}}.
\end{equation}
Substituting a shock radius $r_\mathrm{TS}=9''$ and a radius of the
nebula $r_\mathrm{n}=30''$, corresponding to the inner-torus, we obtain
$\sigma \sim 0.01$, which is twice larger than that reported by
\citet{2002ApJ...569..878G} although they regarded the inner-torus in
Figure \ref{X-ray-image} as a wind termination shock.

In case of the Crab nebula a magnetization parameter of $\sigma=0.003$
based on the expansion velocity was calculated by
\citet{1984ApJ...283..694K},  while \cite{2002phD...Mori.OsakaU} also
reported a larger magnetization parameter of $\sigma \sim 0.03$ based on
imaging analyses utilizing {\itshape Chandra}.

\subsection{South jet}
The south jet in the pulsar vicinity is remarkably different from that
seen in the Crab.  In case of the Crab nebula, the jet seems to flow out
directly from the pulsar, even {\itshape Chandra} cannot resolve the
origin of the jet
\citep{2000ApJ...536L..81W,2002ApJ...577L..49H,2004ApJ...609..186M}.  In
contrast, the jet of PSR B1509$-$58 extends over $\sim10''$ at the
origin.  Also, the jet itself is broad and brighter than its torus.  If
the newly discovered ring feature is a wind termination shock, the south
jet of PSR B1509$-$58 may flow out from the vicinity of the shock front
or from the inner-torus, namely the post shock region.

The appearance of the south jet at the origin seems to match the recent
theoretical scenario in which the hoop stress in the pulsar wind
collimates the poloidal flow \citep{2002MNRAS.329L..34L}.  This idea was
of course introduced for the Crab nebula, however, it is difficult to
explain the narrow jet because the magnetic pinch works well only
downstream of the shock.  Downstream, the magnetic hoop stress is
believed to be able to squeeze the ``mildly relativistic'' flow towards
the rotating axis.  The observed features indicate that this model can
be applied to PSR B1509$-$58.

\citet{2004A&A...421.1063D} reported two dimensional relativistic MHD
simulations of pulsar jets in various configurations.  Interestingly the
observed morphology of the PWN around PSR B1509$-$58 closely resembles
the results of the simulations for $\sigma=0.03$.  They argued that
producing a jet by magnetic hoop stress requires a comparatively large
magnetization parameter, $\sigma \geq 0.01$, which is coincident with
the calculated value shown above.  It is therefore argued that the south
jet of PSR B1509$-$58 originates in the post-shock equatorial flow.
%\newpage

%\section{Conclusion}

%%%%%%%%%%%%%%%%%%%%%%%%%%%%%%%%%%%%%%%

\bibliographystyle{aipproc} \bibliography{refs}

%%%
% See the manual for the detail.
%%%
%\begin{thebibliography}{}
% Journals(e.g. A\&A,ApJ,AJ,NMRAS,PASP ...)
% Authors, Year, Journal, Vol#, Page#
% Journal Title Abbreviation >> http://www.asj.or.jp/pasj/Jabb.html
%\bibitem[Aauthor et al.(2001)]{key-1}
%   Aauthor, A., Bauthor, B., Cauthor, C.\ 2001, PASJ, vol, page
% Books
%\bibitem[Aauthor \& Author(2001a)]{key-2}
%   Aauthor, A., Author, B.\ 2001, Name of Book(Publisher, Tokyo) ch0
% Books
%\bibitem[Aauthor \& Bauthor(2001b)]{key-3}
%  Aauthor, A., Bauthor, B.\ 2001, Name of Book(Publisher, Tokyo) page0
%......
% Editorial Books
%\bibitem[Dauthor(2001)]{key-n}
%  Dauthor A. A.\ 2001, in Name of Book,
%   ed Editor D.\ Editor(Publisher, Tokyo) page0
%\end{thebibliography}

\end{document}